\documentclass[aps,amsfonts,amsmath,prd,preprint,nofootinbib]{revtex4}

\newcommand{\beq}{\begin{equation}}
\newcommand{\eeq}{\end{equation}}

\usepackage{epsfig}

\epsfclipon
\usepackage{epsfig,bm}
\usepackage{amssymb,amsmath}

\begin{document}

\rightline{\footnotesize{CERN-PH-TH/2009-088}}
\rightline{\footnotesize{HD-THEP-09-12}}


\title{{\Large Kahler Moduli Inflation Revisited}}

\vspace{.5cm}

\author{{\large Jose J. Blanco-Pillado$^{a,}$}}
\email{jose@cosmos.phy.tufts.edu}
\author{{\large Duncan Buck$^{b,}$}}
\email{ppxdb@nottingham.ac.uk}
\author{{\large Edmund J. Copeland$^{b,}$}}
\email{ed.copeland@nottingham.ac.uk}
\author{{\large Marta Gomez-Reino$^{c,}$}}
\email{marta.gomez-reino.perez@cern.ch}
\author{{\large Nelson J. Nunes$^{d,}$}}
\email{n.nunes@thphys.uni-heidelberg.de}

\vspace{.4cm}

\affiliation{$^a${\em Institute of Cosmology,
Department of Physics and Astronomy\\ Tufts University, Medford, MA
02155, US}}
\affiliation{$^b${\em School of Physics and Astronomy, University of Nottingham\\
University Park, Nottingham, NG7 2RD, UK}}
\affiliation{$^c${\em Theory Division, Physics Department, CERN,\\
CH-1211, Geneva 23, Switzerland}}
\affiliation{$^d${\em Institute f\"{u}r Theoretische Physik   Philosophenweg 16,\\
 D-69120 Heidelberg, Germany}}

\vspace{.6cm}

\begin{abstract}
We perform a detailed numerical analysis of inflationary solutions in Kahler 
moduli of type IIB flux compactifications. We show that there are
inflationary solutions even when all the fields play an important role
in the overall shape of the scalar potential. Moreover, there exists a
direction of attraction for the inflationary trajectories that
correspond to the constant volume direction. This basin of attraction
enables the system to have an island of stability in the set of
initial conditions. We provide explicit examples of these
trajectories, compute the corresponding tilt of the density
perturbations power spectrum and show that they provide a robust 
prediction of $n_s \approx 0.96$ for 60 e-folds of inflation. 
\end{abstract}

\maketitle

\section{Introduction}


The theory of inflation has been very successful in resolving many
of the most important puzzles in early universe cosmology. However, 
there is, at the moment, no compelling evidence as to what could 
actually produce this period of accelerated expansion. It is therefore
interesting to look for ways to understand this period of cosmic
evolution within the framework provided by a fundamental theory.
String theory is, at present, one of the most promising candidates
for a fundamental theory and has inspired many attempts to embed
inflation within it (for reviews see \cite{Lidsey:1999mc,Quevedo:2002xw,
Linde:2005dd,Cline:2006hu,Kallosh:2007ig,Burgess:2007pz, McAllister:2007bg}). 
One of the most important challenges that has faced string
phenomenology for a long time has been the issue of moduli 
stabilization \cite{Brustein:1992nk,Barreiro:1998aj,GKP,KKLT}. Any 
successful model of low energy string theory should somehow be able 
to fix all the moduli such that it would be compatible with our 
current observations. On the other hand, the universe is not a static 
place but dynamical, so one is also interested in learning how we 
reached this low energy state, making other regions of the moduli 
space, and not only the final minimum, important in order to 
confront theory with cosmological observations. Taking this into 
account it is not surprising that recent developments of general 
methods of moduli stabilisation \cite{GKP}, and in particular 
stabilisation techniques of KKLT \cite{KKLT}, have led to a large 
number of new inflationary scenarios using either the open string 
moduli related to the position of a mobile D-brane
\cite{Brane-Inflation} or the closed string moduli coming from 
the compactification \cite{Modular-Inflation,Kahler-Inflation,Roulette-Inflation} 
as the relevant scalar fields. 

This plethora of models should not be taken as a sign that inflation 
is easy to achieve within string theory. In fact, it is probably 
safe to say that it is just the opposite since most of these models 
have some degree of fine tuning in them. Indeed some of these 
problems were already encountered in the early models of modular
inflation \cite{Binetruy:1986ss, Brustein:1992nk, Banks:1995dp}. 
The main reason for these difficulties is the fact that
the majority of these models are based on ${\cal N}=1$ supergravity
theories that have notorious problems to overcome if they are to 
satisfy the slow roll conditions necessary to have a successful 
inflationary model, the so-called $\eta$ problem \cite{eta-problem}. 
It is therefore very interesting to look for models based within string 
theory that can somehow alleviate or ameliorate these difficulties.

In this paper we will focus on a particular model of modular inflation
that makes use of the special form of the potential for the
Kahler moduli \cite{Kahler-Inflation} enabling it to avoid the $\eta$-problem.
The model is embedded within the Large Volume scenario developed in 
\cite{Large-Volume} something that, as we will show, turns
out to be an important ingredient for the arguments presented in
\cite{Kahler-Inflation}. These Large Volume Models have been extensively
studied in the last few years, due to their phenomenological
interest as an explicit example within string theory of the large
extra dimensional scenarios envisioned by \cite{ADD}. It is therefore
very interesting to study the cosmological implications of these
type of models since they could provide us with a way to select
the correct properties of the compactification scenario that we
would like to have. 

The purpose of this work is two fold. Firstly we 
demonstrate that there are inflationary solutions consistent with
current observational data even when all of the moduli fields are 
allowed to vary during the cosmological evolution. Secondly, 
we show with explicit examples that the set 
of initial conditions that lead to a stable evolution, i.e., that 
avoid a runaway in the decompactification direction, is fairly wide. 
This property results from the existence of a basin of attraction in 
field space. There is an overlap in places between our work and that 
of Bond et al. \cite{Roulette-Inflation}, and where appropriate we will 
compare our results with theirs. 

The outline of this paper is as follows. In Section 2 we introduce the 
models under study. In Section 3 we briefly review the mechanism of 
Kahler moduli inflation. In Section 4 we numerically investigate the
parameter space of the model where one can obtain inflationary
trajectories and illustrate our results through some examples. 
In Section 5 we study the basin of attraction of the inflationary 
solutions, before concluding in Section 6.

\section{The Kahler moduli potential}

Our inflationary scenario can be obtained within a class of
Type IIB flux compactification models on a Calabi-Yau orientifold. In
this context it has been shown in \cite{GKP,KKLT} that the
superpotentials generated by background fluxes and by non-perturbative 
effects like instantons or gaugino condensation may generate a scalar 
potential that stabilizes all the geometric moduli coming from the 
compactification. More concretely, the introduction of background
fluxes in the model induces a superpotential that freezes the dilaton 
as well as the complex structure moduli to their values at their 
supersymmetric minimum \cite{GKP}. The remaining moduli, that is, 
the Kahler moduli, could be then stabilized by non-perturbative
contributions to the superpotential \cite{KKLT}. The resulting
effective 4D description of the Kahler moduli $T_i$ is an ${\cal N}=1$ 
supergravity theory with a superpotential of the type,
\begin{equation}
W=W_0+\sum_{i=2}^nA_ie^{-a_i T_i} \,.
\end{equation}
In this formula $W_0$ is the perturbative contribution coming from the
fluxes, which depends only on the frozen dilaton and the complex
structure moduli, and therefore we will take to be a constant. There
is also a non-perturbative piece depending on the Kahler moduli 
$T_i$ where $A_i$ and $a_i$ are model dependent constants. 

The $F$-term scalar potential is then given by the standard ${\cal N}=1$ formula
\begin{equation}\label{pot}
V(T_i)=e^{{\cal{K}}}[{\cal K}^{i\bar{\jmath}}D_iW D_{\bar \jmath}\bar W-3|W|^2] \,,
\end{equation}
where $D_iW = \partial_iW +(\partial_i{\cal{K}})W$ is the covariant
derivative of the superpotential and ${\cal K}$ is the Kahler
potential for $T_i$. In this paper we will concentrate in the kind of type 
IIB models presented in \cite{Large-Volume} in which the 
$\alpha'$ corrections to the potential are taken into account. For
these type IIB models the expression for the $\alpha'$-corrected
Kahler potential is given by \cite{BBHL}
\begin{equation}
{\cal{K}}_{\alpha\prime}= - 2 \ln \left({\cal{V}}+\frac{\xi}{2}\right) \,,
\end{equation}
where ${\cal V}$ denotes the overall volume of the Calabi-Yau manifold
in string units and $\xi = - {{\zeta(3) \chi(M)}\over {2 (2 \pi)^3}}$ is
proportional to $\zeta(3) \approx 1.2$. The Euler characteristic of
the compactification manifold $M$ is given by $\chi(M) = 2(h^{(1,1)} -
h^{(1,2)})$ where $h^{(1,1)}$ and $h^{(1,2)}$ are the Hodge numbers of
the Calabi-Yau. We will concentrate on models for which $\xi>0$ 
(or equivalently with more complex structure moduli than Kahler
moduli, $h^{(1,2)}>h^{(1,1)}$). As was explained in
\cite{BB,Large-Volume}, the reason for this is that in order to have 
the non-supersymmetric minimum at large volume the leading
contribution to the scalar potential coming from the 
$\alpha'$ correction should be positive.

Following \cite{Kahler-Inflation} we will consider models for which 
the internal volume of the Calabi-Yau can be written in the form,
\begin{eqnarray}\label{vol}
{\cal{V}}=\frac{\alpha}{2\sqrt{2}} \,
\left[(T_1+\bar{T}_1)^{\frac{3}{2}}-
\sum_{i=2}^n{\lambda_i(T_i+\bar{T}_i)^{\frac{3}{2}}} \right] 
= \alpha \left( \tau_1^{3/2} - \sum_{i = 2}^n \lambda_i \tau_i^{3/2} \right) \,,
\end{eqnarray}
where the complex Kahler moduli are given by $T_i=\tau_i+i\theta_i$, 
with $\tau_i$ describing the volume of the internal four cycles
present in the Calabi-Yau and $\theta_i$ are their corresponding 
axionic partners. The parameters $\alpha$ and $\lambda_i$ are model 
dependent constants that can be computed once we have identified a 
particular Calabi-Yau. These models correspond to compactifications 
for which only the diagonal intersection numbers of the Calabi-Yau 
are non-vanishing.

Taking into account the form of the Kahler function one can then
easily compute the Kahler metric for an arbitrary number of moduli, namely,
{\small\begin{eqnarray}
\label{eqn:generalkahlermetric}
&&{\cal K}_{1\bar{1}} = \frac{3\alpha^{4/3} (4 {\cal V} -\xi + 6\alpha 
\sum_{k=2}^n \lambda_k \tau_k^{3/2})}{4(2{\cal V}+\xi)^2 ({\cal V} +
  \alpha \sum_{k=2} \lambda_k \tau_k^{3/2})^{1/3}} \,, \quad\,\,\,\,\,\,
{\cal K}_{i\bar{\jmath}}=\frac{9\alpha^2\lambda_i\lambda_j
\sqrt{\tau_i}\sqrt{\tau_j}}{2(2{\cal{V}}+\xi)^2}\,, \\
&&{\cal K}_{1\bar{\jmath}} = - \frac{9\alpha^{5/3} \lambda_j
  \sqrt{\tau_j} ({\cal V} + \alpha \sum_{k=2}^n \lambda_k 
\tau_k^{3/2})^{1/3}}{2(2{\cal V} + \xi)^2} \,, \quad
{\cal K}_{i\bar{\imath}}=
\frac{3\alpha\lambda_i(2{\cal{V}}+\xi+6\alpha\lambda_i
\tau_i^{3/2})}{4 (2{\cal{V}}+\xi)^2\sqrt{\tau_i}} \,,
\end{eqnarray}}
which can be inverted to give,
{\small\begin{eqnarray}
\label{eqn:lvolKij}
\nonumber
{\cal K}^{1\bar{1}}&=&\frac{4(2{\cal{V}}+\xi)({\cal{V}}+\alpha\sum_{k=2}^n
\lambda_k\tau_k^{3/2})^{1/3}(2{\cal{V}}+\xi+6\alpha(\sum_{k=2}^n\lambda_k\tau_k^{3/2}))
}{3\alpha^{4/3}(4{\cal{V}}-\xi)} \,, \quad 
{\cal K}^{i\bar{\jmath}}=\frac{8(2{\cal{V}}+\xi)\tau_i\tau_j}{4{\cal{V}}-\xi} \,,
\\
{\cal K}^{1\bar{\jmath}}&=&\frac{8(2{\cal{V}}+\xi)\tau_j({\cal{V}}
+\alpha \sum_{k=2}^n\lambda_k\tau_k^{3/2})^{2/3}}{\alpha^{2/3}(4{\cal{V}}-\xi)} \,,
\quad 
{\cal K}^{i\bar{\imath}}=\frac{4(2{\cal{V}}+\xi)\sqrt{\tau_i}(4{\cal{V}}-\xi+6\alpha
\lambda_i\tau_i^{3/2})}{3\alpha(4{\cal{V}}-\xi)\lambda_i} \,,
\end{eqnarray}}
where we have rewritten for later convenience $\tau_1$ in terms of ${\cal{V}}$ and $\tau_i$,
$i=2 \ldots n$. With all this information we can use (\ref{pot}) to obtain the F-term scalar 
potential for the moduli fields which we find to be,
{\small\begin{eqnarray}
\label{general-potential}
\hspace{-.4cm}{\mbox{\large $V$}}\!\!&=&\sum_{\substack{i,j = 2 \\ 
i<j}}^n\frac{A_iA_j\cos(a_i\theta_i-a_j\theta_j)}{(4{\cal{V}}-\xi)
(2{\cal{V}}+\xi)^2}e^{-(a_i\tau_i+a_j\tau_j)}\left(32(2{\cal{V}}+\xi)(a_i\tau_i+a_j\tau_j 
+ 2a_ia_j\tau_i\tau_j)+24\xi\right)\nonumber\\ 
\hspace{-.4cm}&+&\frac{12W_0^2\xi}{(4{\cal{V}}-\xi)(2{\cal{V}}+\xi)^2}
+\sum_{i=2}^n \left[ \frac{12e^{-2a_i\tau_i}\xi
  A_i^2}{(4{\cal{V}}-\xi)(2{\cal{V}}
+\xi)^2}+\frac{16(a_iA_i)^2\sqrt{\tau_i}e^{-2a_i\tau_i}}{3\alpha\lambda_i
(2{\cal{V}}+\xi)} \right.\\ \nonumber
\hspace{-.4cm}&+&
\left. \frac{32e^{-2a_i\tau_i}a_iA_i^2\tau_i(1+a_i\tau_i)}
{(4{\cal{V}}-\xi)(2{\cal{V}}+\xi)}
+\frac{8W_0A_ie^{-a_i\tau_i}\cos(a_i\theta_i)}{(4{\cal{V}}-\xi)(2{\cal{V}}+\xi)}
\left(\frac{3\xi}{2{\cal{V}}+\xi}+4a_i\tau_i\right) \right] + V_{uplift} \,.
\end{eqnarray}}
In this expression for the potential we have introduced an additional
uplift term of the form $V_{uplift}$. The purpose of this term is to
uplift the minima of the potential from an anti-de Sitter minimum to a 
nearly Minkowski vacuum. Its origin in model building has been the
subject of some debate. It could be achieved by breaking explicitly 
supersymmetry through the introduction of anti-branes located in a
region with strong red-shift, as suggested in \cite{KKLT}, or in other
alternative ways involving vector multiplets
\cite{BKQ,Achucarro:2006zf}. Also, from a low-energy effective field 
theory point of view, it can in principle be implemented by using as 
an uplifting sector any kind of theory leading to spontaneous
supersymmetry breaking, provided the supersymmetric sector is
appropriately shielded from this uplifting sector \cite{grs}. Of
course, the different ways that a term of this form can appear in 
the low energy description of the theory may lead to slightly different 
dependencies on the internal volume. For simplicity, as well as for
the sake of comparison, we will take the same form as it was 
previously assumed in \cite{Kahler-Inflation}, namely 
$V_{uplift}=\frac{\beta}{{\cal{V}}^2}$. Nevertheless it is interesting 
to point out that in these Large Volume Models the presence of an 
uplifting sector is not necessary in order to break supersymmetry 
as a non-supersymmetric minimum is already present at large volume 
\cite{Large-Volume,BB}. This is in fact our case here as well.


\section{Single Field Kahler Moduli Inflation}

In ref.~\cite{Kahler-Inflation} the authors argued that
the scalar potential given by the expression (\ref{general-potential}) 
should be able to support a period of slow roll inflation without 
any fine tuning, making it a natural candidate to realize the idea 
of modular inflation. In this section we will briefly review their 
argument, and in the following sections we will proceed to test how 
general can this argument be made.

The first thing one should take into account is that the form of 
the potential (\ref{general-potential}) simplifies substantially in 
the limit in which ${\cal{V}}\gg 1 $. In this limit, as can be inferred from 
(\ref{vol}), there should be one four-cycle (the one given by $\tau_1$) much 
bigger that the rest, $\tau_1\gg\tau_i$, $i=2,\cdots,n$. Taking this 
into consideration one can approximate the full potential by the expression:
\begin{eqnarray}
\label{eqn:largevolpot}
 V_{LARGE}&=&\sum_{i=2}^n
 \frac{8(a_iA_i)^2}{3 \alpha\lambda_i{\cal{V}}}\sqrt{\tau_i}e^{-2a_i\tau_i}
+\sum_{i=2}^n\frac{4W_0a_iA_i}{{\cal{V}}^2}\tau_ie^{-a_i\tau_i}\cos{(a_i \theta_i)}
+\frac{3\xi W_0^2}{4{\cal{V}}^3}+\frac{\beta}{{\cal{V}}^2} \,,
\end{eqnarray}
where we have only included the leading terms up to ${\cal{O}}(\frac{1}{{\cal{V}}^3})$. 

The basic idea now to have inflation in this model is to look for the
possibility of having a flat enough potential by displacing one of the
fields from its minimum value while keeping the others fixed at their
global minimum values. It is reasonable to expect that this strategy
would lead to a successful inflationary period since the potential
is exponentially suppressed along the directions 
$\tau_i$ ($i=2,\cdots,n$). On the other hand, the authors in 
\cite{Kahler-Inflation} also point out correctly that for this idea to
work one should show that the whole inflationary evolution occurs
along a single $\tau_i$ direction, otherwise one would not be able 
to draw conclusions by looking at that particular slice of the 
potential in field space.

The way they propose to enforce this constraint is the following: 
let us assume for concreteness that inflation happens along the 
$\tau_2$ direction. Then in the limit in which $a_i\tau_i \gg1$, 
for $i=2,\cdots,n$, the authors of \cite{Kahler-Inflation} claim that 
by imposing that the parameters appearing in the potential satisfy 
the condition $\rho \ll 1$, where
\begin{equation}
\label{eqn:stabcondition}
 \rho \equiv \frac{\lambda_2}{a_2^{3/2}}:
\sum_{i=2}^{n}\frac{\lambda_i}{a_i^{3/2}}~,
\end{equation}
the minimum of the potential
along the other field directions remain virtually unchanged even if
one displaces $\tau_2$ from its global minimum value. In other words,
for small enough values of $\rho$, there exists a valley of the
potential very much aligned with the direction of $\tau_2$ and
therefore one can assume that moving along that valley all the fields
except $\tau_2$ would stay in their global minimum.

Assuming that this is the case, one can then proceed to approximate
the potential along the inflaton direction $\tau_2$ as,
\begin{equation}
\label{V2}
V_{LARGE} = \frac{B W_0^2}{{\cal{V}}^3} - \frac{4 W_0 a_2 A_2 \tau_2 
e^{-a_2\tau_2}}{{\cal{V}}^2} \,,
\end{equation}
where $B$ includes several terms from Eq. (\ref{eqn:largevolpot})
that depend on the parameters of the potential as well as on the
values of the other fields at their minimum. Also note that 
the axions $\theta_i$ have been set to their minimum, for which 
$\cos{(a_i \theta_i)}=-1$. This is needed in order for a minimum 
for all the fields $\tau_i$ at finite values to exist. Otherwise 
one would have a runaway behavior for some of them.

We can now obtain the values of the slow roll parameters for this
potential at large values of $\tau_2$ by using
their conventional definitions in the single field inflation models, 
namely (we work in Planck units $M_P=1$),
\begin{equation}
\label{eqn:epsilon}
\epsilon = \frac{1}{2}\left(\frac{V'}{V}\right)^2\,,\hspace{1cm}
\eta =  \left(\frac{V''}{V}\right) \,,
\end{equation}
where the primes denote derivatives with respect to the canonically
normalised field $\psi$, defined by normalising the kinetic term for
the inflaton. In the single field inflation approximation we discuss here
and to leading order in the volume we see that,
\begin{equation}
\psi = \sqrt{\frac{4\alpha\lambda_2}{3{\cal{V}}}}\tau_2^{3/4} \,,
\end{equation}
which in turn means that the slow roll parameters are given by  \cite{Kahler-Inflation},
\begin{eqnarray}
\label{eqn:epsdetail}
&&\epsilon =  \frac{32 {\cal{V}}^3} {3\alpha B^2 \lambda_2 W_0^2} a_2^2
A_2^2 \sqrt{\tau_2}(1-a_2\tau_2)^2e^{-2a_2\tau_2} \,, \\
\label{eqn:etadetail}
&&\eta = -\frac{4a_2A_2 {\cal{V}}^2 }{3\alpha \lambda_2\sqrt{\tau_2}B
W_0}(1-9a_2\tau_2+4(a_2\tau_2)^2)e^{-a_2\tau_2} \,,
\end{eqnarray}
and in the limit of slow roll, the associated scalar spectral index 
and tensor to scalar ratio $r$ are given by 
\begin{eqnarray}
\label{spec-index}
n_s-1 &=& 2\eta - 6\epsilon  \,, \\
r &\sim& 12.4 \epsilon \,.
\end{eqnarray}

The number of e-foldings can be computed within this approximate
potential by,
\begin{eqnarray}
\label{eqn:Nefold1}
N_e &=& \int_{\psi_{end}}^{\psi} \frac{V}{V'}d\psi \approx
\frac{-3 B W_0 \alpha \lambda_2 }{16 a_2 A_2 {\cal{V}}^2}
\int_{\tau_2^{end}}^{\tau_2} \frac{e^{a_2 \tau_2}}{\sqrt{\tau_2} (1-a_2\tau_2)} d\tau_2 \,,
\end{eqnarray}
where $\tau_2^{end}$ is taken to be the point in field space where
the slow roll conditions break down i.e. when $\epsilon = \eta =
{\cal{O}}(1)$. It is clear from the expressions 
(\ref{eqn:epsdetail})--(\ref{eqn:Nefold1}) that one 
can get small enough slow-roll parameters as well as a large number 
of e-folds, just by starting at large enough values of $\tau_2$ so 
that ${\cal{V}}^2 e^{-a_2 \tau_2}\ll 1$. Taking into account that 
we are in the slow roll regime, we can then calculate the amplitude 
of the adiabatic scalar perturbations using the expression,
\begin{equation}
\label{amplitude}
P = {1\over {150 \pi^2}} \left({{V}\over {\epsilon}}\right) \simeq 
{1\over {150 \pi^2}}  \left({{3 \alpha B^3 W_0^4 
\lambda_2e^{2 a_2\tau_2}}\over {32
    {\cal V}^6 a_2^2 A_2^2{\sqrt{\tau_2} (1- a_2 \tau_2)^2}}}\right)
\end{equation}
In \cite{Kahler-Inflation}, the authors proposed a `footprint' for 
their model of Kahler inflation. Normalising the density perturbations 
to COBE and seeking $N_e$ efoldings of inflation (typically between
50-60) they obtained the results
\begin{eqnarray}
\eta &\simeq& - {1\over N_e} \,, \hspace{1.4cm}
\epsilon < 10^{-12} \,, \\
 \label{n-range} 0.960 &<& n_s < 0.967 \,, \hspace{.4cm}
0 < |r| < 10^{-10}  \,.
\end{eqnarray}
Such a small value for $\epsilon$ at horizon exit implies that the 
inflationary energy scale is of order 
$V_{\rm inf} \sim 10^{13} {\rm GeV}$, which in turn implies that
tensor modes would be unobservable. A final point that they make is 
that for the model to work, the internal volume ${\cal V}$ is found 
numerically to live within a range of values 
\begin{equation}
\label{vol-range}
10^5 l_s^6 \leq {\cal{V}} \leq 10^7  l_s^6 \,,
\end{equation}
where $l_s = (2\pi) \sqrt{\alpha'}$. It is remarkable how narrow the 
range of $n_s$ is in Eq.~(\ref{n-range}) and how relatively restrictive 
the range of allowed volumes are Eq.~(\ref{vol-range}). One of the 
goals of this work will be to see whether these footprints really do 
define the model when we allow for the volume modulus and other moduli 
fields to evolve. 

\section{Full Kahler moduli Inflation}

The discussion in the previous section suggests that inflation may be
naturally realized in a large subset of string compactifications. This
is an interesting claim so we would like to carefully study the
validity of the approximations made as well as compare the observable
quantities estimated earlier, such as the number of e-folds, the
validity of the assumption $\rho \ll 1$, the constancy of the volume 
modulus and the scalar spectral index, with the more accurate results 
obtained by numerical integration of the full equations of motion
using the full potential in (\ref{general-potential}) instead of the 
approximate large volume one in (\ref{V2}). Our approach differs in 
detail from that adopted by Bond et al. \cite{Roulette-Inflation} 
in that we will be allowing for a number of the moduli fields to 
vary, including the volume modulus. This will allow us to fully 
explore the validity of the assumption that the volume remains 
effectively constant during inflation. In their approach, the volume 
modulus was kept constant and an analysis of the region of parameter 
space which led to inflation was based upon that assumption. 

\subsection{Numerical evolution}

The equations of motion for our moduli fields can be obtained by
varying the minimal $N=1$, $d=4$ effective SUGRA action of the form 
(in Planck units),
\begin{equation}
S=-\int{d^{4}x\sqrt{-g}\left(\frac{1}{2}R+{\cal K}_{i \bar \jmath}\partial_\mu
T^i\partial^\mu\bar{T}^{\bar{\jmath}}+V(T^m,\bar T^{\bar m})\right)}
\end{equation}
where ${\cal K}_{i\bar{\jmath}}$ is the Kahler metric ,
$T^i$ and $\bar{T}^{\bar{\jmath}}$ are the complex chiral
fields. Considering a
spatially flat FRW spacetime we get,
\begin{eqnarray}
\label{eqn:modulieom}
&&  \ddot{T}^l +
  3H\dot{T}^l+
  \Gamma^l_{ij}\dot{T}^i\dot{T}^j + {\cal K}^{l\bar{k}}\partial_{\bar{k}}V=0\\
&&  3H^2=\left({\cal K}_{i\bar{\jmath}}
\dot{T}^i\dot{\bar{T}}^{\bar{\jmath}}+V\right)\nonumber
\end{eqnarray}
where we have used the definition of the connections of the
Kahler metric $ \Gamma^l_{ij} =  {\cal K}^{l\bar{k}}\frac{\partial
{\cal K}_{i\bar{k}}}{{{\partial{T}^{{j}}}}}$. 
Armed with the full equations of motion 
we can now explore numerically the evolution of all the fields 
and find out what regions of moduli space are suitable for inflation. 

\subsection{Example 1}

Following \cite{Kahler-Inflation} we first analyze the case
where the parameters are such that ${\cal V}\gg 1$ and $\rho \ll 1$ and we
only displace the inflaton ($\tau_2$) from its global minimum
value. Our numerical integration confirms the predictions of the
previous analytic arguments. We observe that all the other fields
remain nearly constant during the whole evolution while $\tau_2$
slowly rolls down to its minimum, essentially reproducing the single
field scenario discussed earlier.

An example with these properties can be obtained by taking
the following set of parameters,
\begin{eqnarray}
\label{values}
    \xi &=& 24, \quad \alpha = 1, \quad \lambda_2 = {1\over 100}, \quad \lambda_3 =
    1, \quad a_2 = 20 \pi, \quad a_3 = {\pi \over 2} \nonumber \\
\quad A_2&=&{1\over 300}, \quad A_3={1\over 300}, \quad \beta =
1.984002914\times 10^{-6}, \quad W_0 = 2~.
\end{eqnarray}
We have chosen a viable example of this scenario with the minimal 
number of fields possible, which is three. We first obtain the global minimum 
of the potential, i.e. the minimum at zero cosmological
constant, finding it to be at,
\begin{eqnarray}
\label{globalminimum-1}
\tau^f_1 = 35189.343156992, \quad \tau^f_2 &=& 0.302053449, 
\quad \tau^f_3 = 5.886085128, \nonumber \\
 \quad {\cal V}^f &=& 6.601 \times 10^6 ~.
\end{eqnarray}
We see that indeed this is a large volume compactification
scenario so we should be well within the regime of applicability
of the approximations that we indicated in the previous section.
On the other hand, we have chosen these parameters to have
\begin{equation}
\rho \approx 10^{-5},
\end{equation}
so we expect that the value of the volume at the minimum
should remain pretty much unaffected by the displacement of $\tau_2$. 
We choose the initial value of the inflaton to be 
$\tau^i_2 = 0.8510534498$ and find numerically the new values of the 
local minima in the $\tau_1$ and $\tau_3$ directions for this case to be,
\begin{eqnarray}
\label{localminimum-1}
\tau^i_1 = 35244.7673818281,\quad \tau^i_3 = 5.887497350,
\quad {\cal  V}^i =6.616 \times 10^6  \,.
\end{eqnarray}
Comparing these values to ones obtained in the global
minimum one can clearly see that the displacement of $\tau_2$ 
does not have a big impact on the position of the local minima 
for the other fields in agreement with the analytic arguments given above.

We have chosen this particular value of $\tau^i_2$ to illustrate 
that it is straightforward to obtain sixty e-folds of inflation 
with this set of values. Similarly we have normalized the parameters 
in the potential namely, $A_2$, $A_3$ and $W_0$ in (\ref{values}) 
so that we obtain the correct magnitude of the perturbations 
for this particular solution.

We can now compute the observational signatures of this model 
within the analytic approximations described above. Using the 
expressions given in (\ref{V2}), (\ref{eqn:epsdetail}), 
(\ref{eqn:etadetail}), (\ref{eqn:Nefold1}) and (\ref{amplitude}) 
we find,\footnote{Note that for this set of
parameters one should take $B \approx 0.002$ and $\tau_2 \approx \tau_2^i$.}
\begin{eqnarray}
\label{analytic-results}
N_e &\simeq& 61 \,,\,\,\,V_{inf} = 10^{13}~GeV \,,\,\,\, P = 4\times 10^{-10}\,, \\
\epsilon &=& 4\times 10^{-17} \,,\,\,\, \eta = - 0.0165 \,,\,\,\, n_s = 0.967 \,.\nonumber
\end{eqnarray}

Having identified a particular set of parameters that leads to a
successful inflationary scenario within the approximations described in
\cite{Kahler-Inflation} we would now like to numerically investigate
this example in detail to confirm its analytic predictions.
We have evolved the system of equations presented in 
(\ref{eqn:modulieom}) considering the complete potential
(\ref{general-potential}), in other words, without using any of the 
approximations we discussed earlier. This way we check that the fields 
behave as we expect them to do all the way to their global minimum even in the
region of the potential that is not well approximated by the analytic
expressions given above. We show in Fig.~1 the last period of the
numerical evolution for $\tau_2$ that starts at $(\tau^i_1,\tau^i_2,\tau^i_3)$. 
We only show the $\tau_2$ trajectory since both $\tau_1$ and
$\tau_3$ stay constant throughout the whole evolution until the last 
moment. This confirms that for these set of parameters we can regard the
evolution as effectively a one dimensional problem.

\begin{figure}
\centering\leavevmode
\epsfysize=4cm \epsfbox{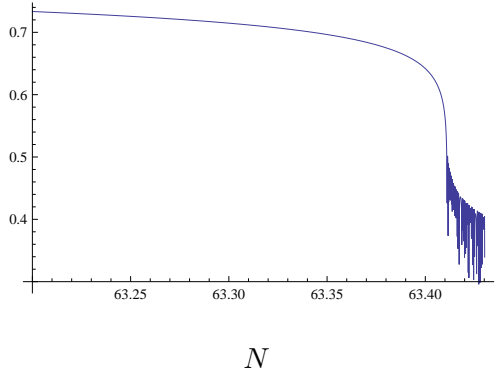}
\hskip -3.5truecm\vskip 0.1truecm{\small $N$} 
\caption[Fig 1] {Evolution of the $\tau_2$ field in the last few
  e-folds in Example 1.}
\end{figure}

Having found the solution numerically we can now calculate
the amplitude of the adiabatic scalar density perturbations directly from
the solutions by computing, 
\begin{equation}
\label{power-spectrum}
P(N) = {1\over {150 \pi^2}} {V(N)\over \epsilon(N)} \,,
\end{equation}
where $N$ denotes the number of e-foldings along the numerical
trajectory and $\epsilon(N)$ correspond to the slow-roll parameter 
which in terms of the Kahler metric and the potential takes the form
\begin{equation}
\epsilon(N) = {{{\cal K}^{i\bar \jmath}\,\nabla_iV\nabla_{\bar \jmath} V}
\over {V^2}} \,.
\end{equation}
We can also extract the spectral index $n_s$ from the expression of the form,
\begin{equation}
\label{spectral-index}
n_s = 1 + {{d\log P(N)}\over {dN}} \,.
\end{equation}
Puting all these expressions together we obtain the following results
numerically,
\begin{eqnarray}
\label{numerical-results}
N_e &\simeq& 63\,,\,\,\, V_{inf} = 10^{13}~GeV  \,,\,\,\, P = 4\times 10^{-10} \,,\\\
\epsilon &=& 4\times 10^{-17} \,, \,\,\, n_s = 0.963\,. \nonumber
\end{eqnarray}
which is in very good agreement with current observational data, and also
with the analytic prediction of Conlon and Quevedo
\cite{Kahler-Inflation} given in this case by  Eqs.~(\ref{analytic-results}). 

\subsection{Example 2}

It is interesting to note that we can still find a successful scenario for
inflation within these type of models even when some of the approximations
used in the previous analytic arguments break down for a particular
set of parameters. Let us consider for example what happens when one 
relaxes the constraint of considering a very large volume. We can
accomplish this by just considering a smaller value of the $W_0$,
namely the following parameters \footnote{Note that from the form of the 
potential in (\ref{eqn:largevolpot}) one can easily check that the value of the volume 
at the minimum is at leading order proportional to $W_0$, while the values 
of the rest of the fields are proportional to $\frac{W_0}{{\cal V}}$ 
(see for example \cite{Large-Volume}). 
This implies in particular that by rescaling the parameter $W_0$ then one 
rescales the value of the volume at the minimum while leaving the 
minimum values of the rest of the moduli invariant.},
\begin{eqnarray}
    \xi &=& 24, \quad \alpha = 1, \quad \lambda_2 = {1\over 100}, \quad \lambda_3 =
    1, \quad a_2 = 20 \pi, \quad a_3 = {\pi \over 2} \nonumber \\
\quad A_2&=&{1\over 300}, \quad A_3={1\over 300}, \quad \beta =
3.29801836\times 10^{-9}, \quad W_0 = {1\over 300}
\end{eqnarray}
where we have also changed the value of $\beta$ to be able to set
the global minimum at zero cosmological constant. In this case the
global minimum becomes,
\begin{eqnarray}
\label{globalminimum-2}
\tau^f_1 = 495.4469043856, \quad \tau^f_2 = 0.302090805, \quad \tau^f_3 =
5.8875322868, \quad {\cal V}^f = 11013.6~.
\end{eqnarray}
which has a much smaller value of the volume than the one obtained
in the previous example and which is not within the range given in 
(\ref{vol-range}). In fact, one can check that in this case some of the
expressions for the large volume limit give a poor approximation 
for the real values, due to the fact that the volume is not sufficiently large. 
Nevertheless one can still displace 
$\tau_2$ without disturbing the values of the other fields
in their minima. In particular, one can show that fixing 
$\tau^i_2 = 0.747090805$ one changes the values 
of the local minimum of the potential along the other directions to,
\begin{eqnarray}
\label{localminimum-2}
\tau^i_1 = 496.227462068,\quad \tau^i_3 = 5.888944614, \quad {\cal V}^i = 11039.2\,.
\end{eqnarray}

\begin{figure}[ht]
\begin{center}
\begin{tabular}{ll}
\\
\hskip -0.5cm
\epsfig{file=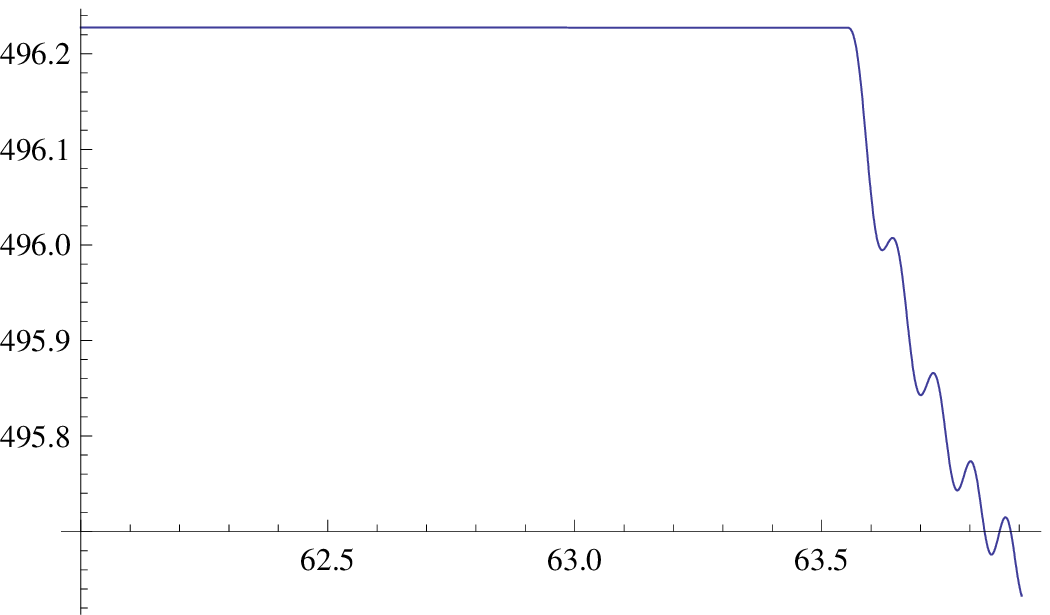, height=4cm}\, \,\, &\,\,\,\,\,\,
\epsfig{file=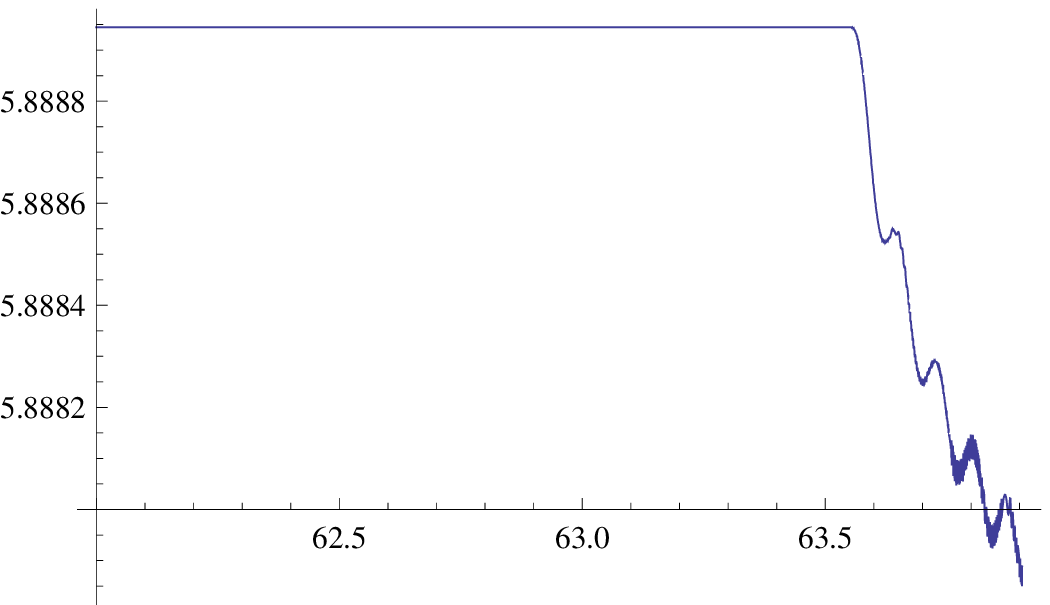, height=4cm}\\
\hskip 2.7truecm{\small $N$}            &
\hskip 3.4truecm{\small $N$}\\
\hskip 3.1truecm{\small (a)}            &
\hskip 3.65truecm{\small (b)}\\
\end{tabular}
\end{center}
\caption{\small{Evolution of the different moduli fields in the last few
  e-folds in Example 2.
a) Evolution of the field $\tau_1$.
b) Evolution of the field $\tau_3$.}}
\label{Ex1}
\end{figure}

We notice that the change in the volume between these two points
in field space is actually much less than $1\% $. In fact, even 
if we increase the value of $\tau^i_2$ considerably the situation 
will not really change, because the value of the volume or of the local 
minimum in which the fields $\tau_1^i$, $\tau_3^i$ sit, remains 
relatively insensitive to the position of $\tau^i_2$.

\begin{figure}[ht]
\begin{center}
\begin{tabular}{ll}
\\
\hskip -0.5cm
\epsfig{file=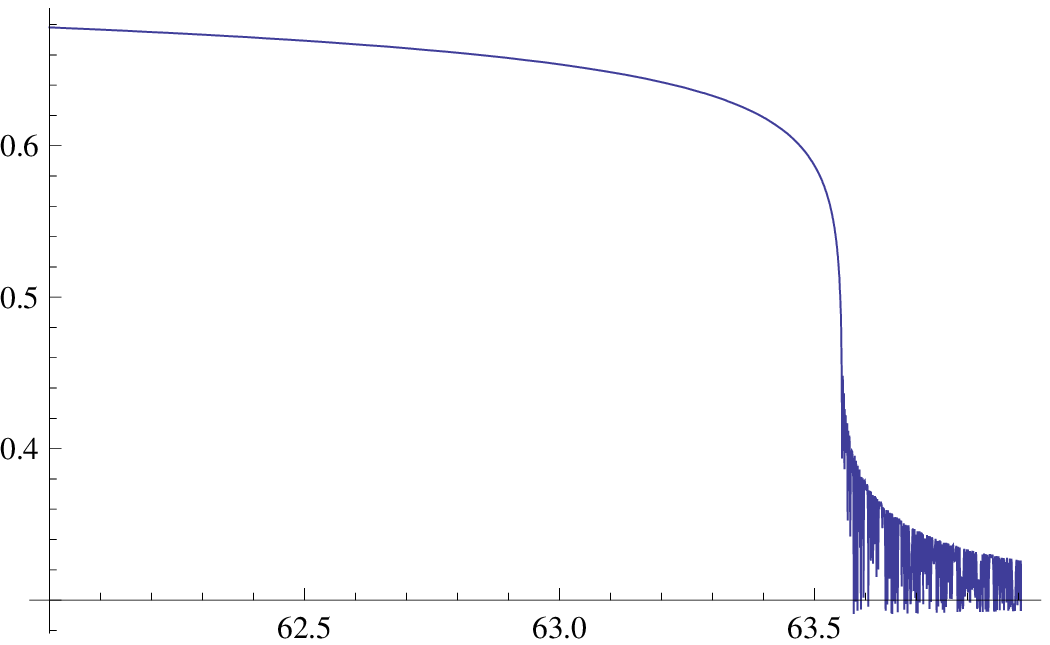, height=4cm}\, \,\, &\,\,\,\,\,\,
\epsfig{file=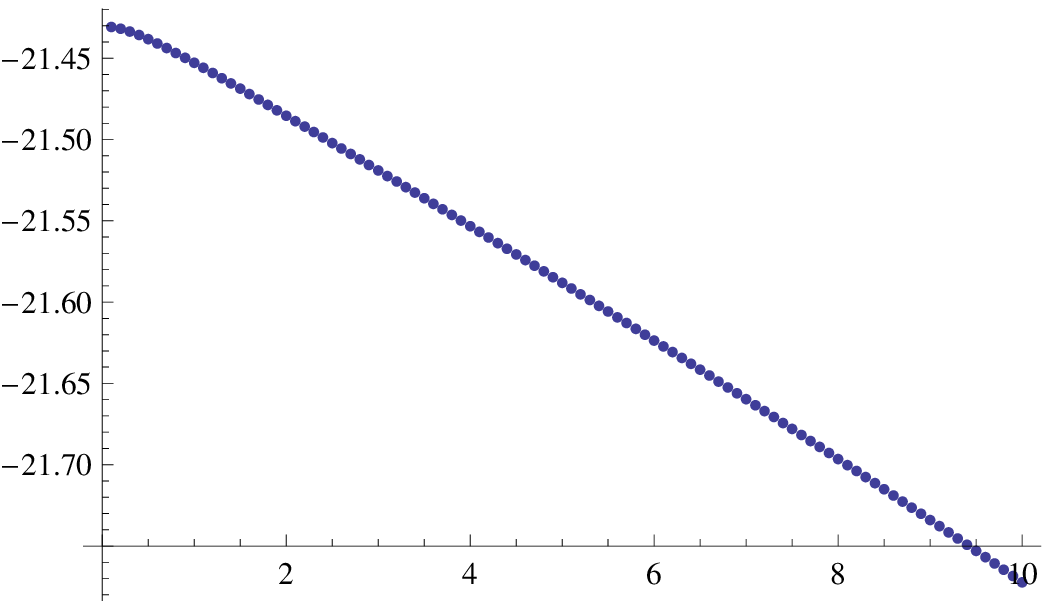, height=4cm}\\
\hskip 2.7truecm{\small $N$}            &
\hskip 3.4truecm{\small $N$}\\
\hskip 3.1truecm{\small (a)}            &
\hskip 3.65truecm{\small (b)}\\
\end{tabular}
\end{center}
\caption{\small{a) Evolution of the moduli field $\tau_2$ (the
    inflaton) in the last few e-folds in Example 2. b) Amplitude 
of the density perturbations in the
10 observationally relevant e-foldings..}}
\label{Ex1-2}
\end{figure}

We show in Figs.~\ref{Ex1}--\ref{Ex1-2} the last few e-folds of 
the numerical evolution that starts at $(\tau^i_1,\tau^i_2,\tau^i_3)$. 
We see that $\tau_1$ and $\tau_3$ stay constant through out the 
whole evolution until the last moment where their values drop 
abruptly to their global minimum values. So effectively our model
is still a one dimensional inflationary model. 

We can now use the expressions given above in Eqs. (\ref{power-spectrum}) and 
(\ref{spectral-index}) to get in this case, 
\begin{equation}
n_s = 0.965
\end{equation}
where we have normalized the potential to obtain the 
correct magnitude of the perturbations within the cosmologically
observable region. (See Fig. \ref{Ex1-2}). 

We conclude from this example that one can extend the region of
the parameter space where a successful inflationary region
can occur even when one can not use some of the large volume
approximations presented in the previous section, but rather the 
expressions computed from the full potential. We will
see in the following that this is also the case for some of the
other assumptions made in \cite{Kahler-Inflation}.

\subsection{Example 3}

As we explained above, the value of the internal volume in the
previous examples remains very much the same during inflation and it is
almost exactly the same as the final value of the volume in the
overall minimum of the potential. This seems to be a stronger
requirement than necessary. In fact, we should only impose that
the volume remains constant during the inflationary period but
it is otherwise free to change substantially after that in its
way to the global minimum. In the following, we will describe one
such example where the volume varies by $45 \%$ from its value
during inflation to the final value.\footnote{There may be other
regions of the parameter space where this change is in fact more
drastic, however, we have restricted ourselves to this milder example for
simplicity.} To illustrate this point let us consider an example with 
the following values of the parameters
\begin{eqnarray}
\label{values2}
    \xi &=& 24, \quad \alpha = 1, \quad \lambda_2 = 1, \quad \lambda_3 =
    1, \quad a_2 = 20 \pi, \quad a_3 = {\pi \over 2} \nonumber \\
\quad A_2&=&{3\over 32}, \quad A_3={1\over 320}, \quad \beta =
6.213734280\times 10^{-9}, \quad W_0 = {1\over 160}
\end{eqnarray}
With these values, $\rho \sim 10^{-3}$, so we are again working in the
regime considered in \cite{Kahler-Inflation}. For this particular example, 
we can show that the global minimum of the potential is located at
\begin{eqnarray}
\label{globalminimum-3}
\tau^f_1 = 751.9457707162, \quad \tau^f_2 =0.2824390994 ,
\quad \tau^f_3 =5.8472434856, \quad {\cal V}^f = 20605.289
\end{eqnarray}
while by displacing the value of $\tau_2^i =0.6624390994$ we
see that the new minimum for the other fields is now found at,
\begin{eqnarray}
\label{localminimum-3}
\tau^i_1 = 970.6098419930, \quad \tau^i_3 =  6.0764936267,
\quad {\cal V}^i =  30170.0176
\end{eqnarray}

\begin{figure}[ht]
\begin{center}
\begin{tabular}{ll}
\\
\hskip -0.5cm
\epsfig{file=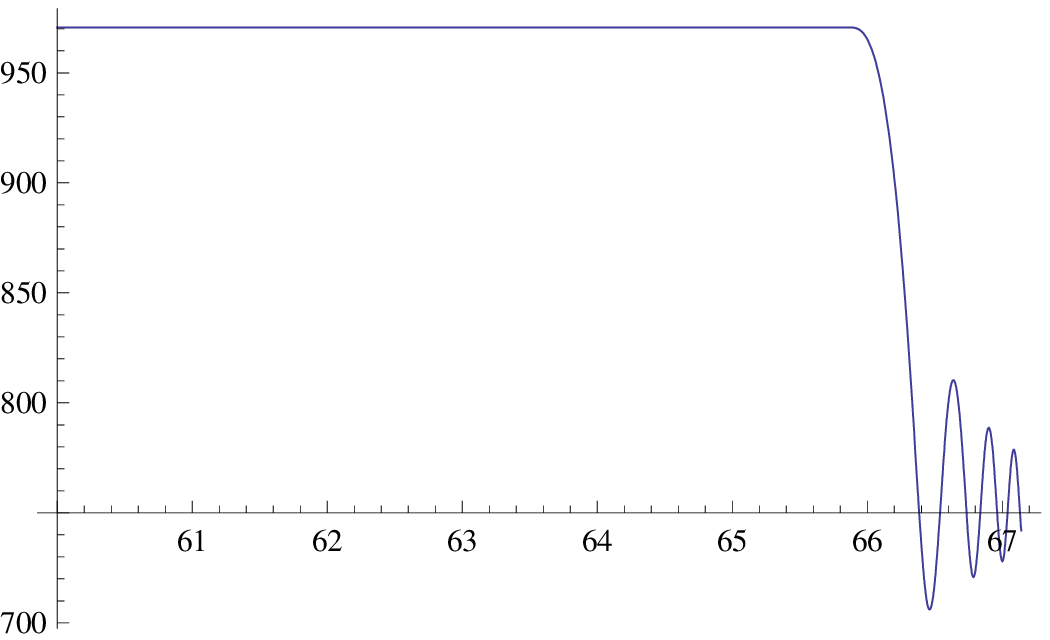, height=4cm}\, \,\, &\,\,\,\,\,\,
\epsfig{file=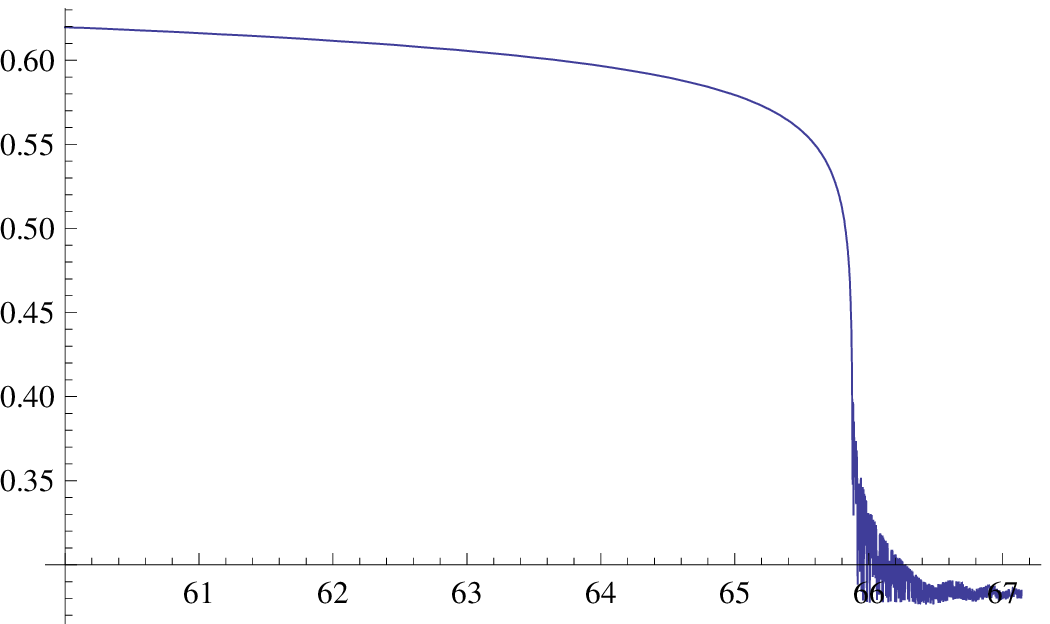, height=4cm}\\
\hskip 2.7truecm{\small $N$}            &
\hskip 3.4truecm{\small $N$}\\
\hskip 3.1truecm{\small (a)}            &
\hskip 3.65truecm{\small (b)}\\
\end{tabular}
\end{center}
\caption{\small{Evolution of the different moduli fields in the last few
  e-folds in Example 3.
a) Evolution of the field $\tau_1$.
b) Evolution of the field $\tau_2$.}}
\label{Ex2}
\end{figure}

\begin{figure}[ht]
\begin{center}
\begin{tabular}{ll}
\\
\hskip -0.5cm
\epsfig{file=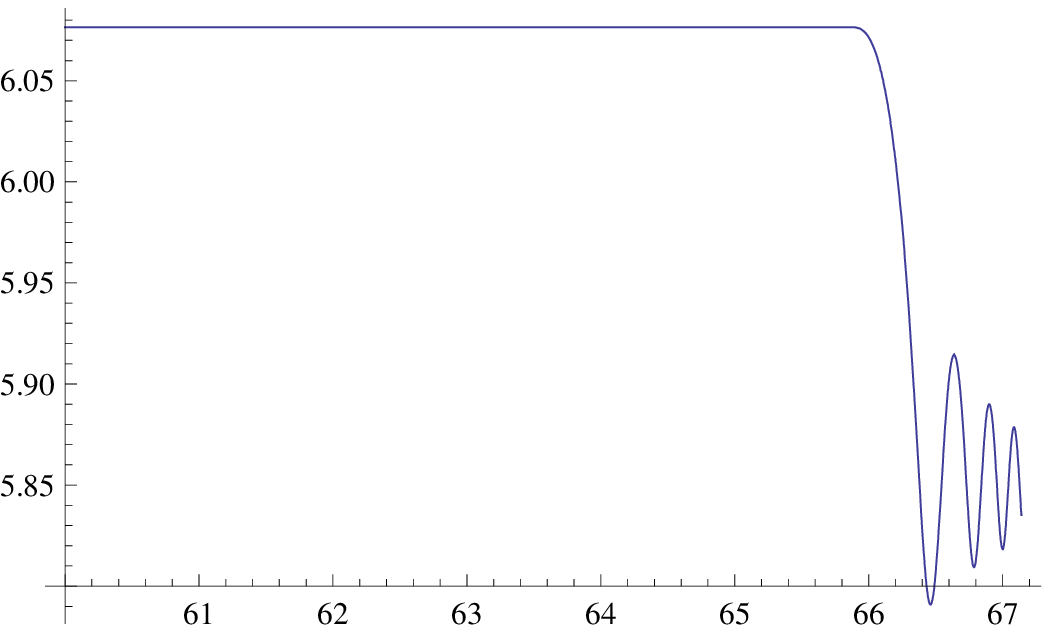, height=4cm}\, \,\, &\,\,\,\,\,\,
\epsfig{file=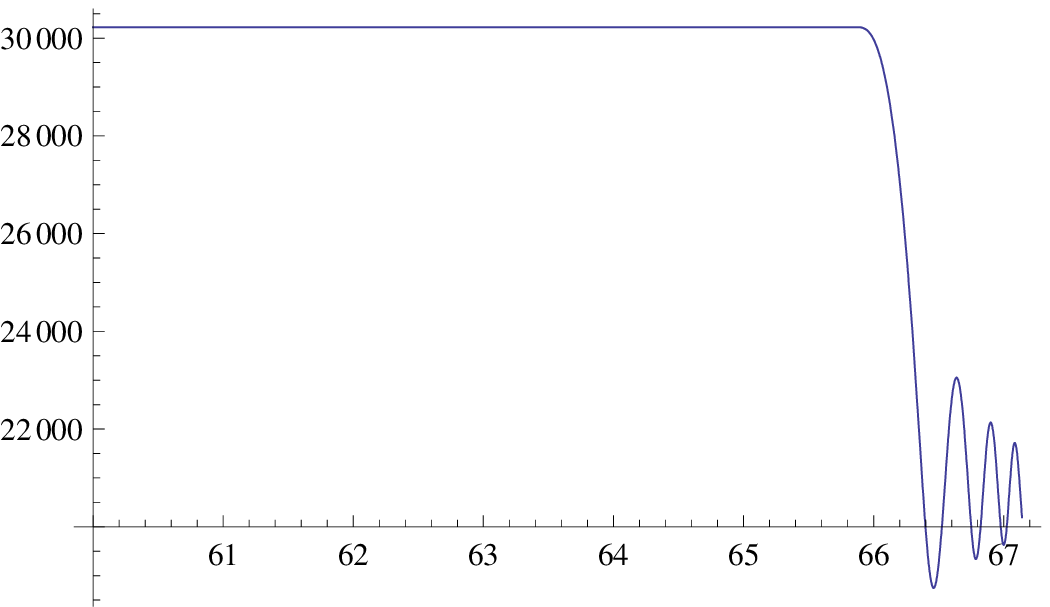, height=4cm}\\
\hskip 2.7truecm{\small $N$}            &
\hskip 3.4truecm{\small $N$}\\
\hskip 3.1truecm{\small (a)}            &
\hskip 3.65truecm{\small (b)}\\
\end{tabular}
\end{center}
\caption{\small{Evolution of the different moduli fields in the last few
  e-folds in Example 3.
a) Evolution of the field $\tau_3$.
b) Evolution of the field ${\cal{V}}$.}}
\label{Ex2-2}
\end{figure}

Once again we evolve the system of equations presented in (\ref{eqn:modulieom}) 
using the complete potential (\ref{general-potential}) and 
check what is the behaviour of the different fields. We plot 
in Fig.~\ref{Ex2} and in part a) of Fig.\ref{Ex2-2} the results for 
evolution of the fields for this example. In part b) of 
Fig.~\ref{Ex2-2} we show the evolution of the internal volume 
in the last few e-folds. We can clearly see there that the volume 
remains constant for the relevant period of inflation and only
changes to its global minimum value within the last e-fold or so.

What can we conclude from this case compared to the previous examples? 
It is clear that although the set of parameters that we have 
used here represents a slightly different behaviour from
the one described in \cite{Kahler-Inflation}, in particular the fact 
that the volume modulus can change quite considerably at the end of 
inflation, nevertheless, it still represents a perfectly valid
inflationary period regarding its observational signatures
so once again this example increases the acceptable region of the
parameter space within this kind of model. Actually we have again 
normalized the parameters in the potential so that we obtain the correct magnitude of the 
perturbations and therefore we can use the expressions 
(\ref{spectral-index}) together with (\ref{power-spectrum}) to 
get in this case
\begin{equation}
n_s = 0.967\, ,
\end{equation}
once again perfectly consistent with the range predicted in
\cite{Kahler-Inflation}.

We can therefore see from this example that the real condition in 
order for a successful period of inflation to take place is that 
the volume remains constant during inflation only, but not 
necessarily during the whole evolution of the fields.
%

\subsection{Example 4}

As we have mentioned, the analytic estimates made in
\cite{Kahler-Inflation} for the spectral index $n_s$, are based on 
the assumption that during inflation $\rho \ll 1$. In this final 
example, we relax that condition, and address whether successful 
inflation still occurs in that situation (recall we are allowing 
all the fields to evolve). For this purpose let us consider the following values of the
parameters,
\begin{eqnarray}
\label{values3}
    \xi &=& {1\over 2}, \quad \alpha = {1\over {9 \sqrt{2}}}, 
\quad \lambda_2 = 10, \quad \lambda_3 = 1, \quad a_2 = 
{{2\pi}\over 30}, \quad a_3 = {{2\pi} \over {3}},
 \nonumber \\ 
A_2&=&{1\over {1.7 \times 10^{6}}},\quad A_3={1\over 425}, \quad \beta = 6.9468131457 \times 10^{-5}, 
\quad W_0 = {40\over 17}~.
\end{eqnarray}
which yields,
\begin{equation}
\rho \sim 0.99 \,.
\end{equation}

\begin{figure}[ht]
\begin{center}
\begin{tabular}{ll}
\\
\hskip -0.5cm
\epsfig{file=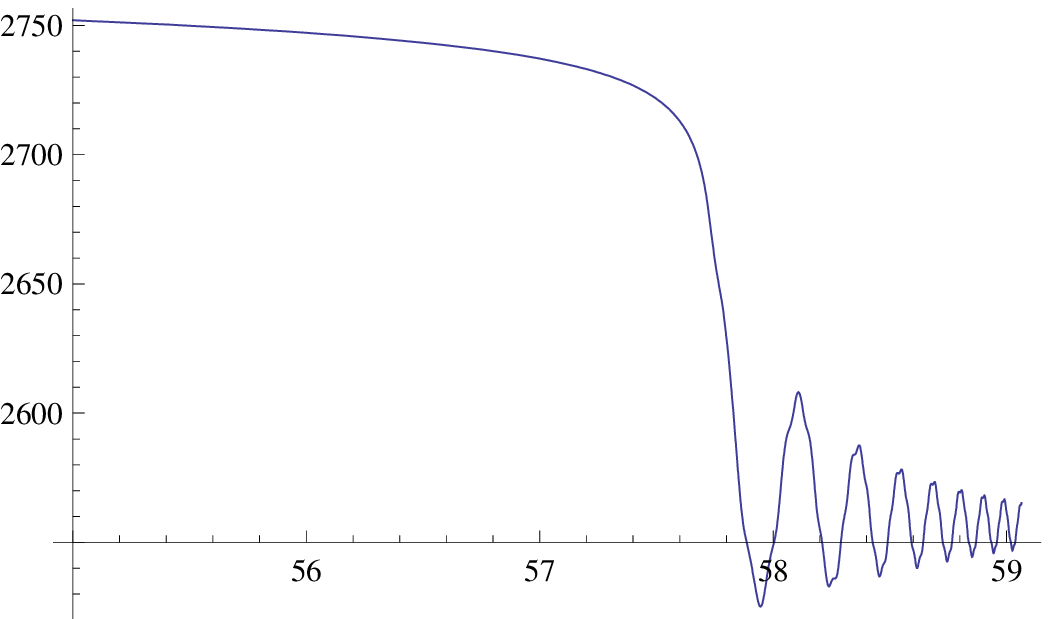, height=4cm}\, \,\, &\,\,\,\,\,\,
\epsfig{file=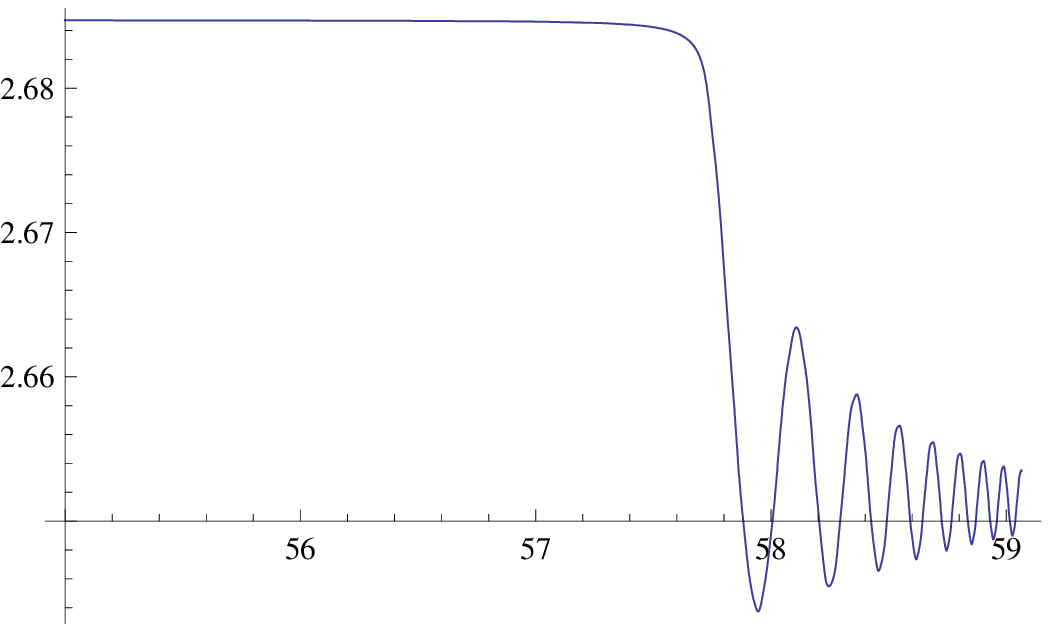, height=4cm}\\
\hskip 2.7truecm{\small $N$}            &
\hskip 3.4truecm{\small $N$}\\
\hskip 3.1truecm{\small (a)}            &
\hskip 3.65truecm{\small (b)}\\
\end{tabular}
\end{center}
\caption{\small{Evolution of the different moduli fields in the last few
  e-folds in Example 4.
a) Evolution of the field $\tau_1$.
b) Evolution of the field $\tau_3$.}}
\label{Ex3}
\end{figure}

\begin{figure}[ht]
\begin{center}
\begin{tabular}{ll}
\\
\hskip -0.5cm
\epsfig{file=EXAMPLE-3-FIG-3.eps, height=4cm}\, \,\, &\,\,\,\,\,\,
\epsfig{file=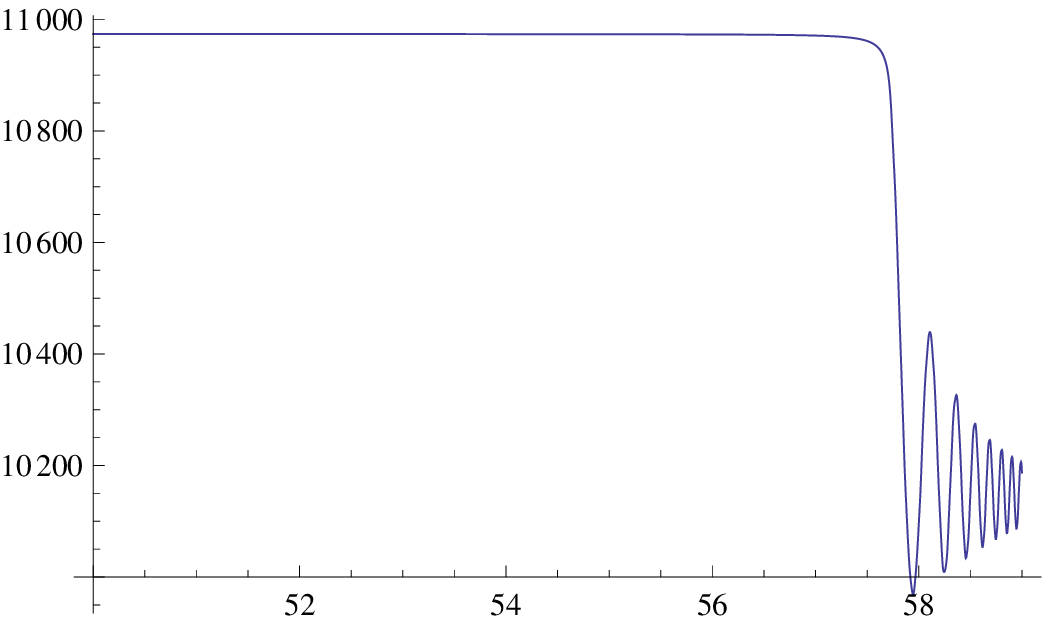, height=4cm}\\
\hskip 2.7truecm{\small $N$}            &
\hskip 3.4truecm{\small $N$}\\
\hskip 3.1truecm{\small (a)}            &
\hskip 3.65truecm{\small (b)}\\
\end{tabular}
\end{center}
\caption{\small{Evolution of the different moduli fields in the last few
  e-folds in Example 4.
a) Evolution of the field $\tau_2$ (the inflaton).
b) Evolution of the field ${\cal V}$.}}
\label{Ex3-2}
\end{figure}

The global minimum of the potential is now located at
\begin{eqnarray}
\label{globalminimum-4}
\tau^f_1 = 2555.95 , \quad \tau^f_2 = 4.7752 ,
\quad \tau^f_3 = 2.6512 , \quad {\cal V}^f = 10143.94363 \,.
\end{eqnarray}
Displacing the value of $\tau_2$ to a substatially larger value,
namely, $\tau_2^i = 78.7752067$ we
see that the new minimum for the remaining fields is found at,
\begin{eqnarray}
\label{localminimum-4}
\tau^i_1 = 2781.185086997 , \quad \tau^i_3 = 2.684717126 ,
\quad {\cal V}^i = 10973.9 \,.
\end{eqnarray}
As in the previous example we can now evolve again the system of equations presented in 
(\ref{eqn:modulieom}) using the complete potential 
(\ref{general-potential}) and check what is the behaviour of 
the different fields. We plot the results in Figs.~\ref{Ex3} and \ref{Ex3-2}. More 
concretely in Fig.~\ref{Ex3} and in part 
a) of Fig.~\ref{Ex3-2} the evolution of the fields $\tau_i$ and in part b) of Fig.~\ref{Ex3-2} 
the evolution of the internal volume in the last few e-folds. 


As before, we find a successful period of inflation with this new  
set of parameters, reproducing the correct amplitude of density 
perturbations and obtaining 
\begin{equation}
n_s = 0.960 \,,
\end{equation}
which is very similar to the range predicted in \cite{Kahler-Inflation}.

The reason why this system of parameters works is that the
initial value of $\tau_2$ is large enough so that the exponential
dependence of the potential with this field makes a negligible
contribution to the calculation of the minima as a function of the 
other two degrees of freedom, namely,
$\tau_3$ and ${\cal V}$. This allows for the possibility of having an
inflationary valley sitting at the minimum of the potential along
those directions, even in cases where $\rho \sim 1$. 

In summary, the examples shown above demonstrate the existence of a 
large region of parameter space within these models with inflationary 
solutions consistent with current cosmological observations even when 
one relaxes most of the constraints stated in \cite{Kahler-Inflation}.


\section{Basin of attraction}

In the previous section, we have considered that the fields $\tau_1$
and $\tau_3$ were initially placed at the local minimum associated
with the displacement of $\tau_2$ from its global minimum, i.e. 
$\tau_1^i = \tau_1^{\rm local}$ and $\tau_3^i = \tau_3^{\rm local}$. 
In this section, we want to relax this assumption and verify whether 
the model allows for some freedom in the choice of initial
conditions, namely we would like to see whether there is a region in the 
space of $\tau_1^i$, $\tau_2^i$ and $\tau_3^i$ that leads to viable
inflationary solutions as good as the ones presented above.

\begin{figure}[ht]
\begin{center}
\begin{tabular}{ll}
\\
\hskip -0.5cm
\epsfig{file=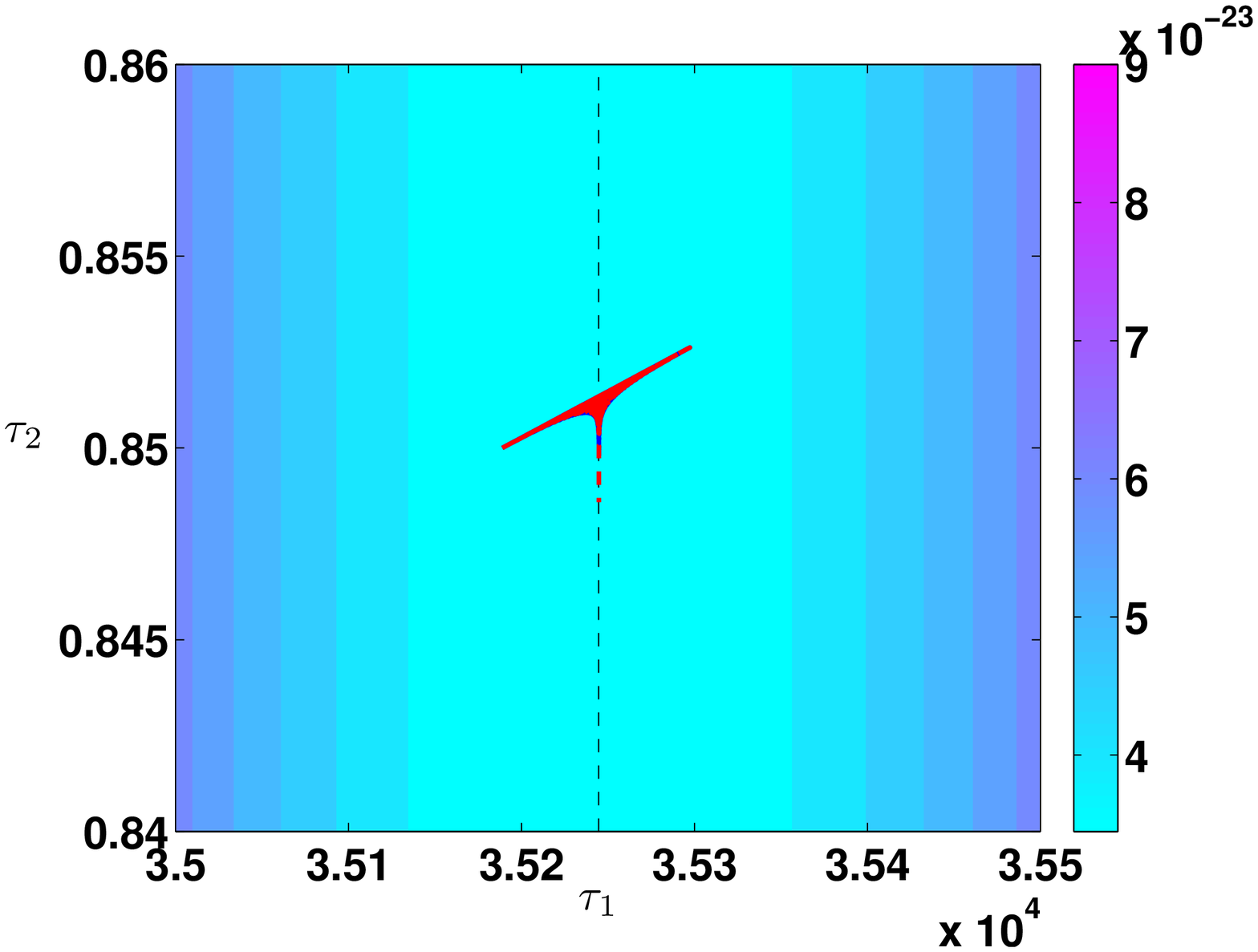, height=5cm}\, \,\, &\,\,\,\,\,\,
\epsfig{file=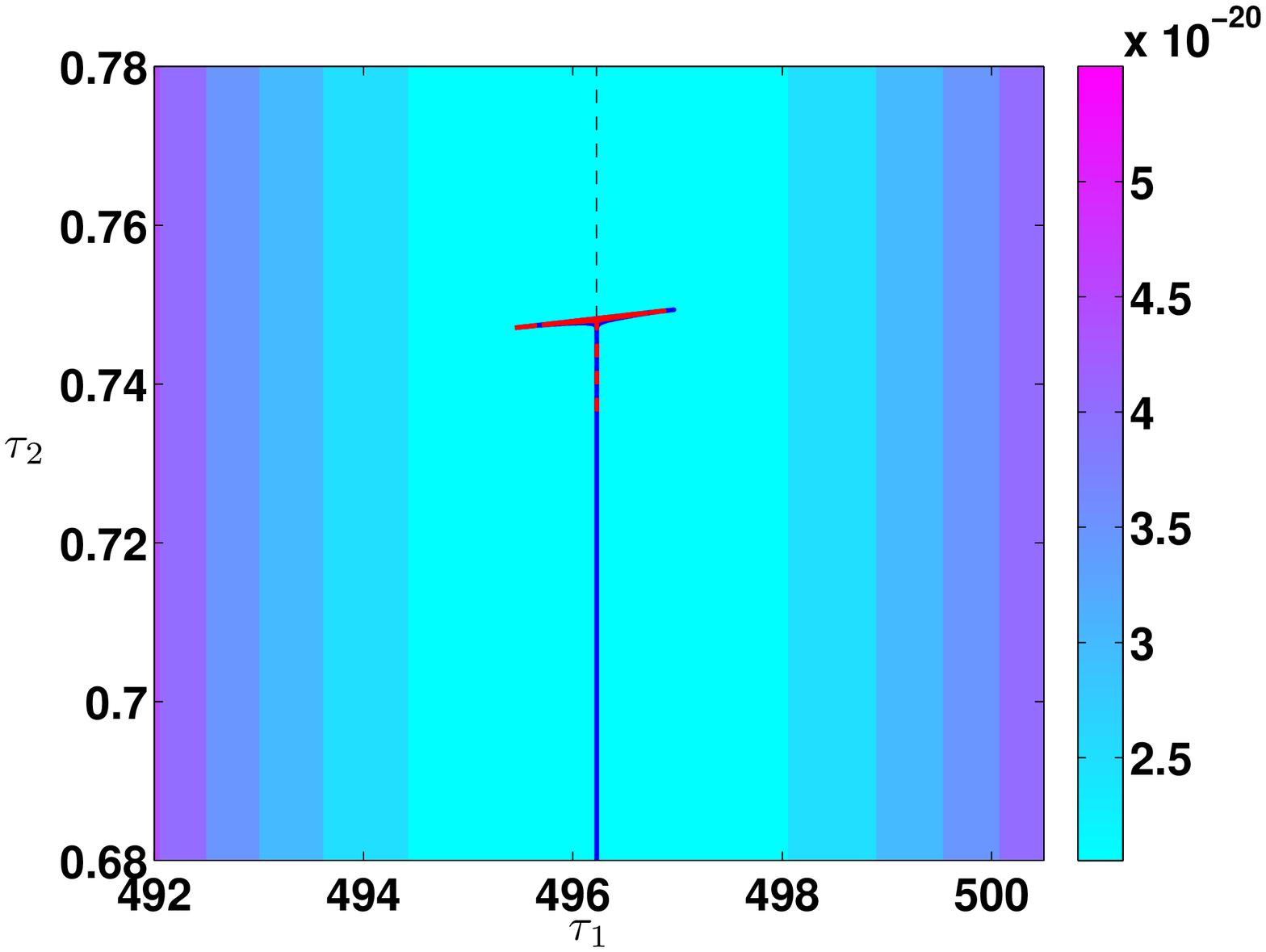, height=5cm}\\
\hskip 3.1truecm{\small (a)}            &
\hskip 3.65truecm{\small (b)}\\
\end{tabular}
\end{center}
\caption{\small{
a) Contour plot of the scalar potential $V$ in the  ($\tau_1,\tau_2$) 
plane for fixed $\tau_3 = \tau_3^{\rm local}$, for example 1. The
dashed line shows the trajectory which maintains volume ${\cal{V}}$ 
constant at this fixed $\tau_3$. b) Contour plot of the scalar
potential $V$ in the  ($\tau_1,\tau_2$) plane for fixed 
$\tau_3 = \tau_3^{\rm local}$, for example 2. The dashed line shows 
the trajectory which maintains volume ${\cal{V}}$ constant at this
fixed $\tau_3$. }--Trajectories are $\theta_2^i=\theta_{2_{min}}$ 
(blue solid line) and  $\theta_2^i\neq\theta_{2_{min}}$ (red dashed line).}
\label{basin}
\end{figure}

\begin{figure}[ht]
\begin{center}
\begin{tabular}{ll}
\\
\hskip -0.5cm
\epsfig{file=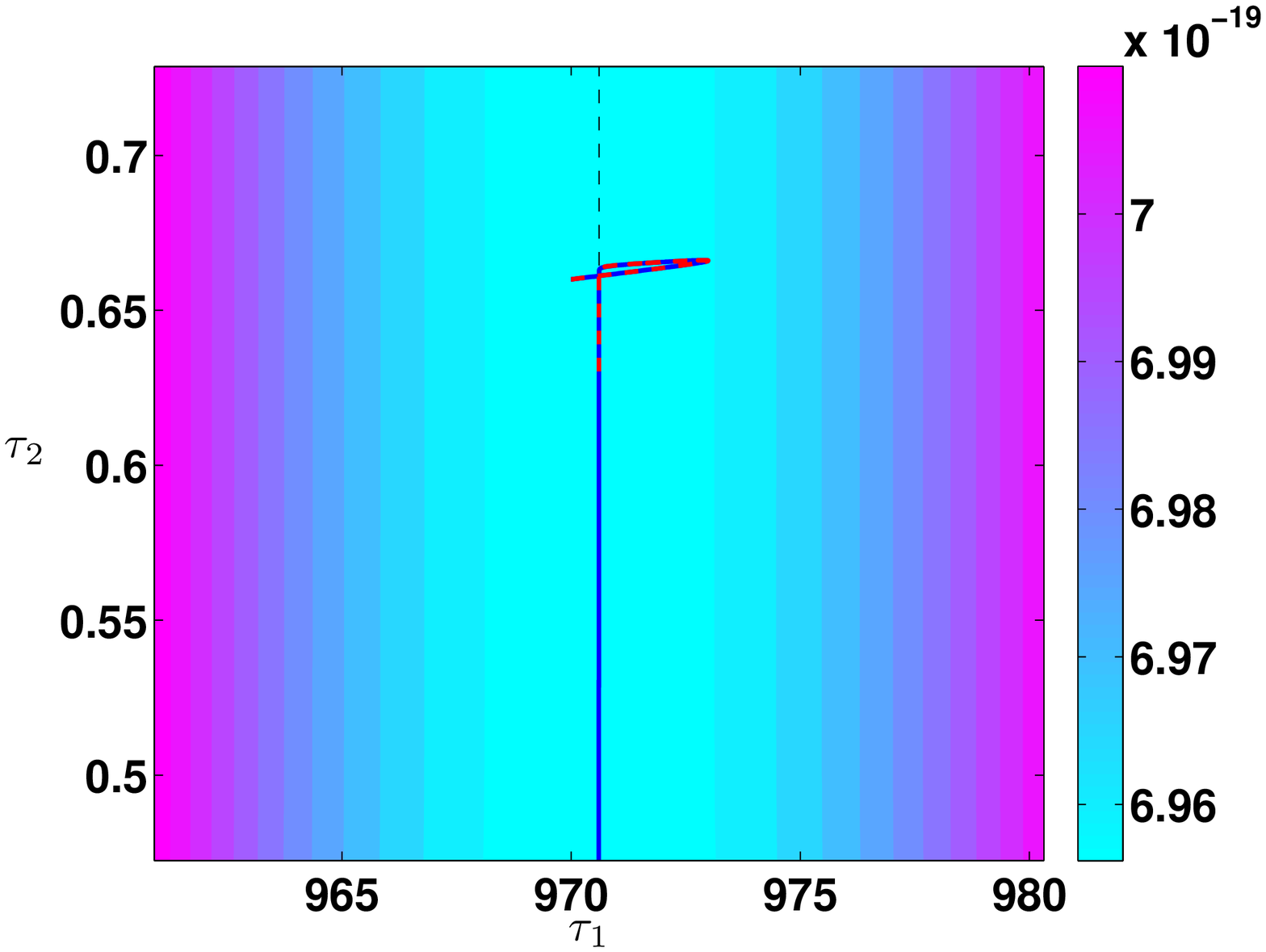, height=5cm}\, \,\, &\,\,\,\,\,\,
\epsfig{file=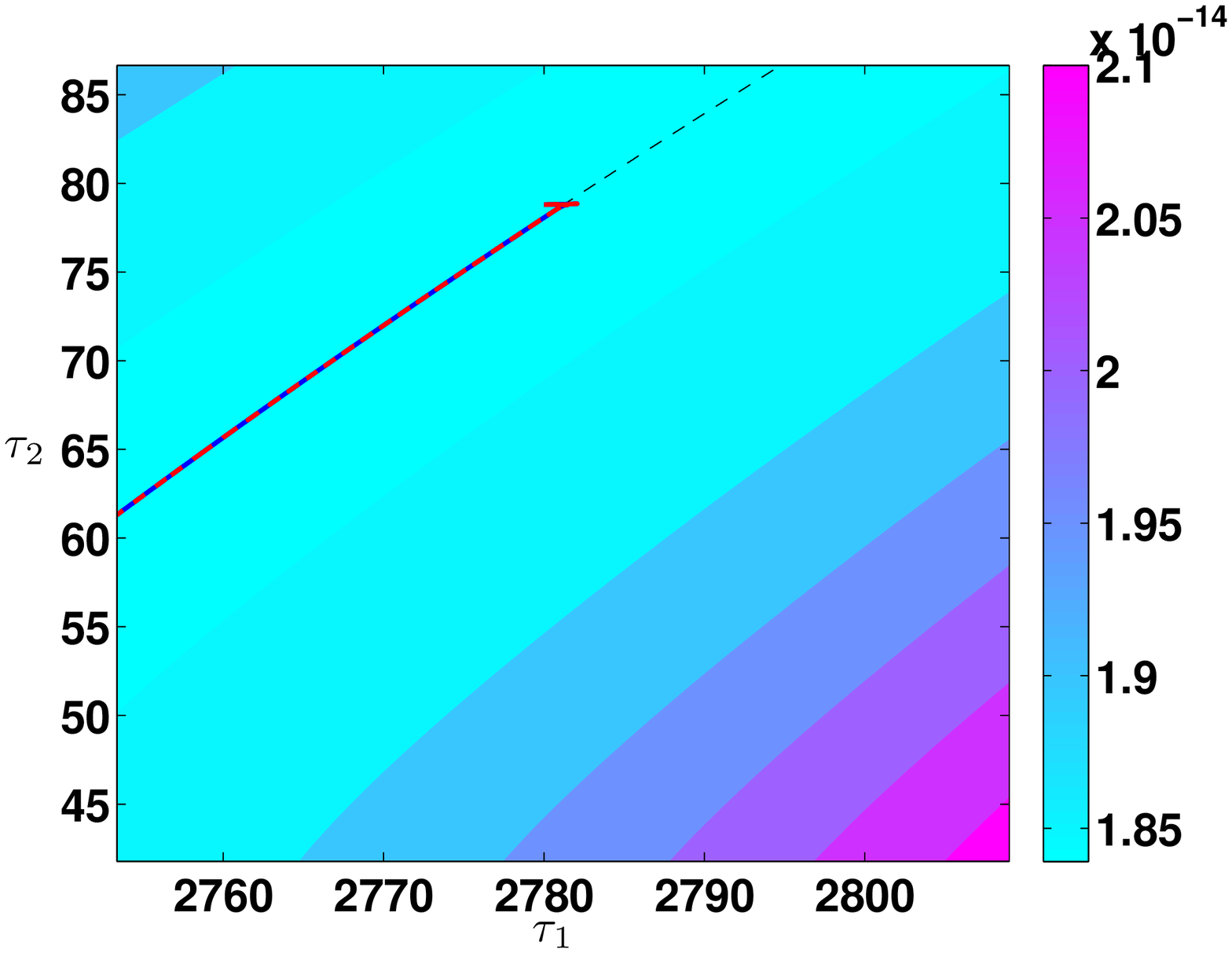, height=5cm}\\
\hskip 3.1truecm{\small (a)}            &
\hskip 3.65truecm{\small (b)}\\
\end{tabular}
\end{center}
\caption{\small{
a) Contour plot of the scalar potential $V$ in the  ($\tau_1,\tau_2$) 
plane for fixed $\tau_3 = \tau_3^{\rm local}$, for example 3. The
dashed line shows the trajectory which maintains volume ${\cal{V}}$ 
constant at this fixed $\tau_3$. b) Contour plot of the scalar
potential $V$ in the  ($\tau_1,\tau_2$) plane for fixed 
$\tau_3 = \tau_3^{\rm local}$, for example 4. The dashed line shows the 
trajectory which maintains 
volume ${\cal{V}}$ constant at this fixed $\tau_3$. In this example, 
the inflationary trajectory is essentially given by a combination of 
two fields. Trajectories are $\theta_2^i=\theta_{2_{min}}$ (blue solid line) 
and  $\theta_2^i\neq\theta_{2_{min}}$ (red dashed line).}}
\label{basin3}
\end{figure}

We note that the relative difference between $\tau_3^{\rm local}$ and 
$\tau_3^f$ as given by Eqns.~(\ref{globalminimum-1}),
(\ref{localminimum-1}), (\ref{globalminimum-2}), (\ref{localminimum-2}), 
(\ref{globalminimum-3}), (\ref{localminimum-3}) and 
(\ref{globalminimum-4}), (\ref{localminimum-4}),  is at the most of 
only of a few percent. Hence, we can consider that for an initial
condition in the vicinity of $\tau_3^{\rm local}$, $\tau_3$ is nearly 
constant during inflation. This simplification allows us to illustrate 
the shape of the scalar potential during inflation, and in particular 
to show that there is a basin of attraction in the ($\tau_1$,$\tau_2$) 
plane. We show this plane in Figs.~\ref{basin}a and \ref{basin}b as
well as Fig.~\ref{basin3}a and \ref{basin3}b corresponding to examples 
1, 2, 3 and 4, respectively. The dashed line represents the direction 
of constant 
volume ${\cal V}$ for fixed $\tau_3^i = \tau_{\rm local}$. We also
show in Figs.~\ref{basin} and \ref{basin3} the full numerical
evolution of the fields with initial conditions slightly away from  
(\ref{localminimum-1}), (\ref{globalminimum-2}), (\ref{localminimum-3}) and
(\ref{localminimum-4}). This choice serves our purpose but initial 
conditions further away from the local minimum are also allowed and 
can in fact increase the number of e-folds of inflation as $\tau_2$ 
can be displaced to higher values.

We see that the basin of attraction not only stabilises the evolution 
of the fields directing them towards the global minimum but also
forces them to satisfy the essential condition ${\cal V} \approx$
constant which is established by the orientation of the basin itself 
in the ($\tau_1,\tau_2$) plane. When inflation terminates, the fields 
quickly evolve to the global minimum and the evolution departs form
the trajectory ${\cal V} \approx$ constant represented in the
figures.

Curiously, this variation of the internal manifold volume ${\cal V}$ 
leads to the existence of two different scales for the gravitino
mass. During inflation ${\cal V} \approx {\cal V}_i$ and once it ends, 
the fields fall to the global minimum where, at least in our  
examples, ${\cal{V}} = {\cal{V}}_f < {\cal{V}}_i$. Given that the gravitino mass is,
\begin{equation}
\label{gravitino-mass}
m_{3/2}^2 \sim e^K|W|^2 \sim W_0^2/({\cal{V}}+\xi/2)^2 \sim W_0^2/{\cal{V}}^2 \,,
\end{equation}
we have ${m_{3/2}}_{\rm inflation}^2 \approx W_0^2/{\cal V}^2_i$ and 
${m_{3/2}}_f^2 = W_0^2/{\cal{V}}^2_f$, leading to a larger gravitino 
mass after inflation. For the sets of parameters in our examples, 
the gravitino masses during inflation are $m_{3/2} = 3\times 10^{-7}$, $3\times 10^{-7}$, 
$ 2 \times 10^{-7}$ and $2 \times 10^{-4}$ in Planck 
units, for examples 1, 2, 3 and 4, respectively. These scales are 
rather high and therefore not very appealing phenomenologically. 
Actually this is typically the case in most of the inflationary 
models built from string theory. This follows from the fact that, as was argued in
\cite{Kallosh:2004yh}, the scale of inflation is generically 
bounded from above by the mass of the gravitino $H\lesssim m_{3/2} $. 
Therefore since these string inflationary models (in order to reproduce the correct 
amplitude for the density perturbations given by current observational data) predict a high
scale of inflation, they also predict as well a high supersymmetry
breaking scale. This is sometimes referred to as the gravitino mass problem. 
This feature is stronger in this class of inflationary models
built from the $\alpha'$-corrected Kahler potential \cite{CY}, where 
the scale of inflation that can be realised within these setups corresponds to  
$H \sim m_{3/2}/{\cal{V}}^{1/2}$ or $H \sim m^{3/2}_{3/2}$ using Eq.~(\ref{gravitino-mass}), 
which will typically give rise to even higher 
supersymmetry breaking scales. Recall however that 
this mass corresponds to the gravitino mass during inflation, which 
does not have to be necessarily the same as the gravitino mass 
in the vacuum. This point has been used for example 
in \cite{Conlon:2008cj,BO} to propose a mechanism which can achieve 
low energy supersymmetry breaking scales, which consists in performing an extra fine-tuning 
in the models so that the gravitino mass during and at the end of inflation are 
substantially different. In this context it is interesting to note that the examples 
mentioned above also display a different gravitino mass during and after
inflation. Unfortunately in our examples we have always found 
$m_{3/2}^i < m_{3/2}^f$ and therefore the gravitino mass at the vacuum 
is heavier than the one during inflation which is going in the wrong 
direction. The question is whether or not  there are trajectories of 
the form described in the previous examples which can lead to the volume 
increasing immediately after inflation. There is no obvious reason 
why this can not happen but it remains a challenge to find an example.

\subsection{Evolving the axions}

We can also investigate what might happen if we allow the axions 
$\theta_i$ to evolve, as well as all the fields $\tau_i$. 
As expected, if the initial conditions for the axions are such 
that they are placed at their minimal values the examples described 
above do not change, as the axions do not get displaced from their 
minimum. However the situation is modified when the initial 
conditions for the axions are such that they are perturbed from 
their minimal values. In such a situation two different scenarios 
emerge. In the case in which only the axion corresponding to the 
inflaton field (that is, the field $\theta_2$) is perturbed, 
we see that the fields evolve in such a way as to reproduce the situation described in 
\cite{Roulette-Inflation} (for the case $\rho \ll 1$). In particular, 
viable inflationary trajectories exist, but the new initial conditions 
allow for a greater variety of trajectories in which the rolling 
of the axion can increase the  number of e-foldings over those 
trajectories restricted to lie only in the $\tau_2$ direction. 
Such evolutions are the red dashed trajectories in Figs.~\ref{basin}, 
\ref{basin3}.   

The second class of scenarios correspond to perturbing an axion which 
belongs to the same multiplet as the field which plays the dominant role in 
the stabilization of the volume, which in our examples would
correspond to the field $\theta_3$. In this case one can show that the set 
of viable initial conditions for inflationary trajectories is restricted to small initial
perturbations in the position of the axion ($\delta \theta_3 \ll 1$) 
away from its minimum. Any other significant perturbation of
$\theta_3$ leads to runaway non-inflationary trajectories. 

The reason for these different behaviors can be understood as follows: 
the rolling of the axion field $\theta_2$ does not have a noticeable 
impact on the position of the minimum of the volume modulus. On the 
other hand, a displacement of the axionic field $\theta_3$ will
instead have an effect on the position of the volume modulus, as can 
be easily read from (\ref{eqn:largevolpot}). This means that almost 
any displacement on the field $\theta_3$ will have the effect of 
displacing the volume modulus from its minimum and as a consequence 
of that the fields would tend to roll towards the decompactification limit.

\section{Conclusions}

Realising inflation in the context of 
string theory has provided a number of difficulties from the onset. 
The former relies on scalar fields slowly evolving in an almost flat
potential, whereas the natural scales for parameters in string 
theory tend to have potentials which are too steep to sustain an 
extended period of inflation.  Moreover, the plethora of moduli fields 
arising in these models makes it difficult to have just one field 
evolving (the inflaton) whilst the others remain fixed in there
minima. Therefore, when a model is proposed which appears to
successfully reconcile these two important disciplines it deserves 
attention. The model proposed by Conlon and Quevedo \cite{Kahler-Inflation} 
is one such example and has been the focus of this work.  

We have performed a detailed numerical analysis of inflationary solutions 
in the Kahler moduli sector of the Large Volume Models built in the 
context of type IIB flux compactifications. Our investigations confirmed 
the key result of \cite{Kahler-Inflation}, namely that there are 
inflationary solutions where all but one of the moduli fields,
(the inflaton), are stabilised to the local minima of the potential. 
We have provided explicit examples of these trajectories, and shown how the corresponding tilt 
of the density perturbations power spectrum leads to a robust
prediction of  $n_s \approx 0.96$ for 60 e-folds of inflation, 
in agreement with the analytic prediction. However, we have gone 
further and showed that even when all the moduli fields play an 
important role in the overall shape of the scalar potential, 
inflationary trajectories still exist. In particular, we have 
demonstrated that there exists a direction of attraction for the 
inflationary trajectories that correspond to the constant volume 
direction. It leads to a basin of attraction which enables the system 
to have an island of stability in the set of initial conditions 
leading to inflation.

 Furthermore we were able to show, using the numerical evolution of the 
fields under the influence of the full potential, that there are still 
successful inflationary trajectories even when one relaxes most of the 
assumptions made in the analytical approximations of 
\cite{Kahler-Inflation}. This is an interesting point that makes the 
conclusions from these type of models much more robust. 
In particular, we considered a case in which the constraint of
considering a very large volume is relaxed, a case where the volume
varies by 45\% when inflation terminates, and a case where $\rho \approx
1$.

Having looked at the evolution of the moduli fields with the axion 
fields restricted to be in their minima, we then extended 
the analysis to allow also the axions to be slightly 
displaced from their equilibrium position. Whereas a variation of the axion $\theta_2$ 
(the partner of the inflaton) still led to a large basin of 
attraction for the inflationary trajectories (as in \cite{Roulette-Inflation}), 
in the case of the axion $\theta_3$ (the one that lives in the same multiplet as 
the moduli field responsible for stabilising the volume modulus), 
a small fluctuation of $\theta_3$ from its true 
minimum value is enough to create runaway solutions where all the fields 
roll towards the decompactification limit. Hence we have a new 
restriction on these class of models, $\theta_3$ needs to be very 
close to its minimum value for inflation to take place.

\section{Acknowledgements}
 We would like to thank Cliff Burgess, Ken Olum, Fernando Quevedo 
and Gianmassimo Tasinato for helpful conversations and Joe Conlon 
for useful discussions during the early stages of this work. 
EJC would like to thank the Royal Society for financial support. 
D.B. acknowledges the STFC for the award of a PhD studentship. N.J.N. 
is supported by Deutsche Forschungsgemeinschaft, TRR33. M.G.-R. and N.J.N. 
would like to thank the Galileo Galilei Institute for 
Theoretical Physics for hospitality and the INFN for partial support 
during the completion of this work. J.J. B.- P. is supported by the 
National Science Foundation under Grant 06533561.

\end{document}